# Can the vertical motions in the eyewall of tropical cyclones support persistent UAV flight?

**Chung-Kiak Poh[1], and Chung-How Poh[2]**

[1] Geographic Information Systems Research Center, Feng Chia University, Taichung 40724, Taiwan

[2] Department of Physics, University of Newcastle, Callaghan, NSW 2308, Australia

Email address: kiak@gis.tw

**Abstract**

Powered flights in the form of manned or unmanned aerial vehicles (UAVs) have been flying into tropical cyclones to obtain vital atmospheric measurements with flight duration typically lasting between 12 and 36 hours. Convective vertical motion properties of tropical cyclones have previously been studied. This work investigates the possibility to achieve persistent flight by harnessing the generally pervasive updrafts in the eyewall of tropical cyclones. A sailplane UAV capable of vertical take-off and landing (VTOL) is proposed and its flight characteristics simulated using the RealFlight® simulator. Results suggest that the concept of persistent flight within the eyewall is promising and may be extendable to the rainband regions.

## 1. Introduction

Tropical cyclone looks serene and elegant from the space above, but the sheer scale of the storm can generate winds in excess of 250 kmh$^{-1}$ and drive storm surge causing massive destruction to the surface below [1,2]. The main structural features of a tropical cyclone are the rainbands, the eye, and the eyewall [3]. The eye has a typical diameter of 32 to 64 km, though its formation mechanism is still not fully understood [3,4]. The eye is the calmest part of the storm with winds that usually do not exceed 24 kmh-1 [3]. The eyewall region consists of a ring of tall thunderstorms that produce heavy rains and often the strongest winds [3].

The wind circulations of a matured tropical cyclone can be broadly divided into the primary and the secondary circulation [5]. The primary circulation refers to the tangential flow rotating about the central axis, and the secondary circulation refers to the "in-up-and-out circulation" (low and middle level inflow, upper-level outflow) [5]. Thus, the general air flow model of a tropical cyclone is air parcels spiraling inwards, upwards and outwards [5].

Despite advances in remote sensing, in-situ measurements with reconnaissance flights are still necessary to obtain accurate location of pressure center and wind speeds to aid reliable forecasting [6]. Of particular interest to forecasters and the public is the maximum sustained 10 m surface wind [7]. UAVs have recently been deployed for tropical cyclone missions. Two such well known UAVs are the NASA Global Hawk and the Aerosonde, with flight endurances of 30 hours and 26 hours (minimal paylaod), respectively [8,9].

Comprehensive study on the vertical motions in intense tropical cyclones has been carried out by Jorgensen et al. [10]. It involved dataset from four mature hurricanes (Anita, David, Frederic and Allen) and a total of 115 penetrations from nine flight sortie at altitudes from 0.5 to 6.1 km [10]. A total of over 3000 updrafts recorded and the downdrafts total was nearly 2000. A convective updraft or downdraft was defined as having vertical velocity ($w$) measurements with at least one point > $|0.5|$ ms$^{-1}$ transversing over at least 500 m of horizontal distance [10]. Stronger cores were further distinguished from the drafts if the $|w|$ > 1 ms$^{-1}$ [10]. Thus, a single draft may contain several cores [10].

Thermal soaring is a form of flight where the flying objects use only convection currents to stay aloft without any additional power source (motor power in the case of airplane, or flapping of wings in the case of birds) [11]. Cross-country flights have been undertaken by birds and man-made sailplanes [11]. The soaring birds gain height by circling in thermals with wing spread until desired height is reached. They then glide forward in a sinking phase to the next available column of thermal [11]. UAVs with autonomous thermal seeking capability have been demonstrated [12,13]. Whenever an updraft is rising faster than the sink rate of a sailplane, there will be a gain of altitude. The rate of climb of a sailplane, $V_c$ with its sink rate, $V_s$ in an ascending updraft of average velocity, $\overline{w}$ is [14]:

$$V_c = \overline{w} - V_s \tag{1}$$

This work investigates the potential to develop persistent UAV by harnessing the updrafts in the eyewall of tropical cyclones. Viability of the concept was assessed based on the work published by Jorgensen et al. on the characteristics of vertical motions in hurricanes [10] and by simulating the performance of a small sailplane UAV. The UAV is intended to advance from an updraft core to the next using similar flight scheme as that of the cross-country flights. The VTOL capability of the UAV is demonstrated by simulation and the effect of payload is also evaluated. The long-term goal of the research is to develop an UAV capable of remaining airborne with the storm throughout its life-span to enable uninterrupted acquisition of dataset, particularly information relating to the 10 m surface wind.

## 2. Methods

### 2.1. Characteristics of vertical motions

A simple graphical representation of the updraft cores within ±5 km from the radius of maximum wind (RMW) was generated using the results reported by Jorgensen et al. [10]. The primary aim was to establish the estimated mean distance between 2 adjacent cores in the eyewall. This information will allow one to assess whether a gliding UAV with a given flight performance has the required range to reach adjacent core, taking into account other possible environmental conditions.

Data analysis by Jorgensen et al. revealed that the size and strength of the drafts and cores were very much similar for all the four hurricanes considered. Four properties considered were the 1) intercepted length of the draft or core, $D$ (loosely refered to as the "diameter"); 2) mean vertical velocity of each draft or core $\bar{w}$; 3) the maximum 1-s vertical velocity in the draft or core $w_{max}$ and 4) total mass transport per unit length normal to the flight track.

The distributions of these properties were found to be approximately linear on a log-normal plot. In both the eyewall and rainband regions updrafts dominated over downdrafts in terms of counts and mass transport. Approximately two to three times more updraft cores than downdraft cores were encountered, particularly at the lower altitudes [10]. 99% of the updraft cores were < 7 km in diameter, and 99% of the downdraft cores were less than about 5.6 km [10]. Area covered by drafts and cores were also studied for an average area size of ~5.9 × $10^4$ km$^2$ centered around the eye. Results suggested that relatively small area of these mature hurricanes was covered by significant vertical motions and that the eyewall region was dominated by updrafts [10]. The dominance was presented as percentages of an annulus defined as ±5 km around the RMW covered by the cores. The inner halve was defined as the eyewall region and the outer region was defined as the rainband region [10]. The altitude of 0.5 km is of particular relevant to surface wind measurement and at such altitude, the percent of eyewall region covered by the updraft and downdraft cores were 54.7% and 22.7%, respectively. Furthermore, for the purpose of graphical representation, the diameter of the updraft cores in the eyewall and rainband regions assumed the top 10% value of 3.2 km at 0.5 km altitude [10].

RMW of Hurricane Anita of 21 km [15] was used in this work to generate the graphical representation of the updraft core distribution at altitude 0.5 km. Given the similarity, it should typify the other three hurricanes or other tropical cyclone in general.

## 2.2. Simulation details

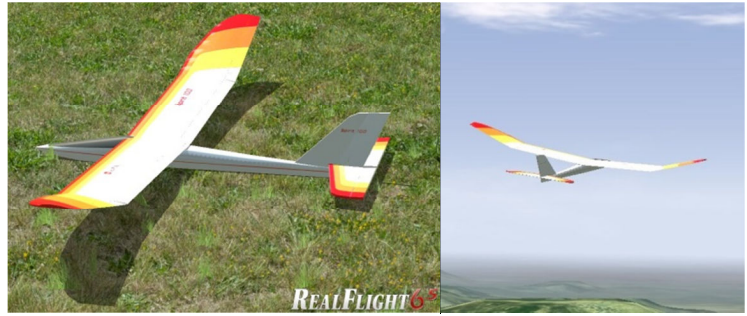

**Fig. 1** The as-supplied simulation model of the GreatPlanes® Spirit 100

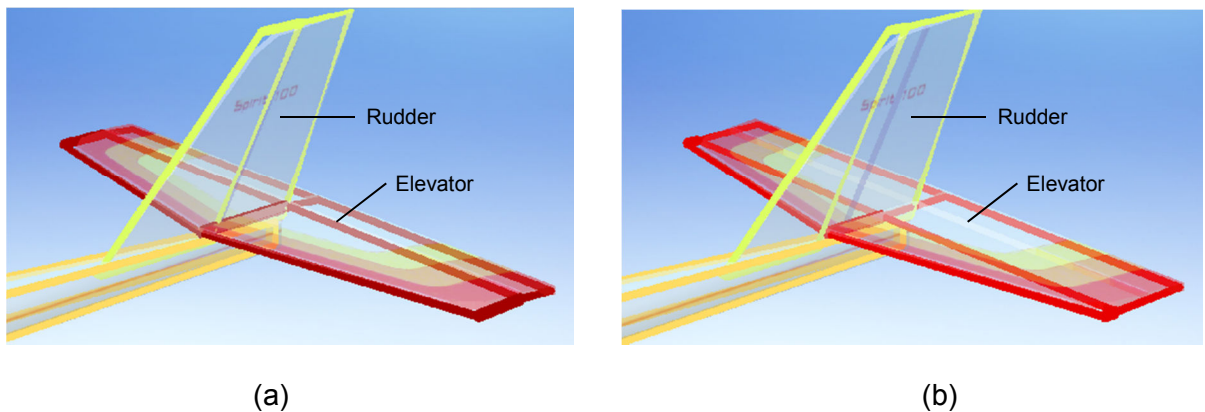

(a)                                   (b)

**Fig. 2** Comparison of rudder and elevator control surfaces on the (a) Spirit 100 and (b) Spirit 100-VT. Generous surface area facilitates VTOL operation.

Simulation work was performed using the RealFlight® 6.5 simulator [16] running on a quad-core 2.2 GHz computer. The Spirit 100 (shown in Fig. 1) with highly-efficient Selig 703 airfoil was used as the base platform to create the gliding UAV. The motorless Spirit 100 has a wing span of 2.53 m and flying weight of 1.7 kg. Modification of the Spirit 100 was done using the Accu-Model™ aircraft editor. Surface areas of the rudder and elevator were increased by adjusting the chord ratios, and the flaps were replaced by a pair of ailerons. Control surface areas of the original Spirit 100 and its modified variant (termed as Spirit 100-VT) are as shown in Fig. 2(a) and (b), respectively. The following components were added to the platform to enable powered flight: A 2204 W brushless motor with 121.72 N thrust, a pair of co-axial counter-rotating folding propellers (size: 432 x 174 mm), and a 6-cell 8000 mAh lithium polymer (Li-po) battery pack. Angular position dependent 3-axis (roll, pitch, yaw) gyro was added to the platform for enhanced flight stability. The completed model weighed in at 3.085 kg.

Gliding performance was characterized using the polar curve, with the motor turned off and the propellers folded. The flight power was acquired by setting the sailplane on a straight and level flight, adjust the throttle until the desired airspeed is reached with the variometer showing zero readout; if the airspeed, altitude and electrical power value remained invariant for 1 minute, equilibrium conditions were considered reached and the power value was recorded as the power requirement. Effect of payload on flight performance was also evaluated at 1, 3, and 5 kg. The center of mass of the aircraft was made to coincide with the center of lift throughout the simulation to ensure flight characteristics were not adversely affected.

## 3. Results and discussion

### 3.1. Properties of updraft cores in the eyewall

Fig. 3 shows a simple graphical representation of the updraft cores within the eyewall and rainband at altitude 0.5 km based on published data mentioned in section 2.1. The diameter of cores in both regions were 3.2 km. The density of updraft in the rainband was included in Fig. 3 as well even though the focus of this study was on the eyewall region. The average core-to-core distances in the eyewall and rainband were estimated to be 2.9 and 4.75 km, respectively.

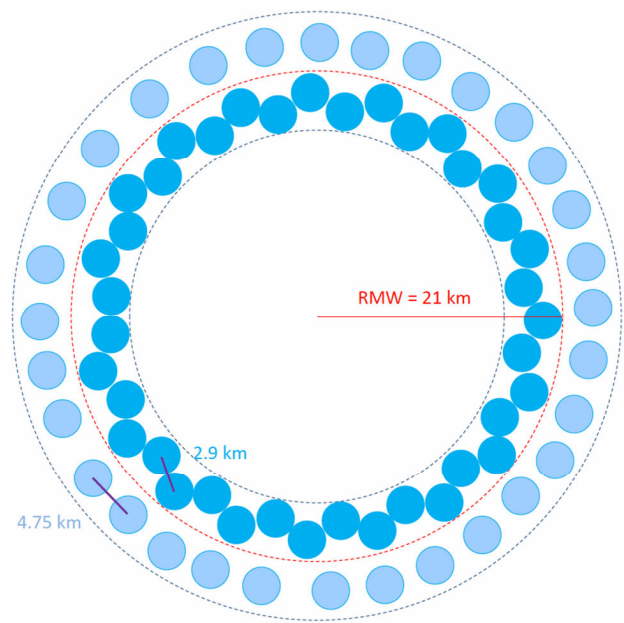

**Fig. 3** A simple graphical model showing a plausible distribution of updraft cores +/- 5 km from the radius of maximum wind (RMW) for Hurricane Anita at altitude 0.5 km.

### 3.2. VTOL capabilities

Normal flight and VTOL capabilities were successfully demonstrated with good control over the roll, pitch and yaw, even with the payload of 5 kg. The use of the counter-rotating propellers was essential as it reduced the propeller torque effect and the P-factor to an almost negligible level, and consequently, virtually no aileron deflection was needed to maintain roll-free vertical hovering. This allowed for smaller ailerons to be used, symmetrical roll-rate on both directions, and improved the hovering stability. Less aileron deflection also resulted in greater vertical acceleration, which can be used to lift heavier payload.

Figs. 4(a) to (f) show the vertical landing sequence for the gliding UAV with 5 kg payload onboard. The landing sequence begun with the final approach and followed by the vertical climb and reduction in airspeed. As the nose pitched upward, the motor thrust was gradually increased until the platform came to a transition hovering (variometer showing 0 ms$^{-1}$). The platform was then brought into a constant rate of descent (-0.9 ms$^{-1}$, in this case) and terminated with the final hovering. Optionally, the platform could be steered with precision onto a landing station, if necessary [Fig. 4(f)].

Fig. 5 shows the hovering power for payload ranging from 0 to 5 kg, and a linear fit has been applied to the plot with a $R^2$ value of 0.9974. The 3.085 kg platform itself consumed 343.68 W for hovering and a linear fitting gave 230 W kg$^{-1}$ for any addition of payload.

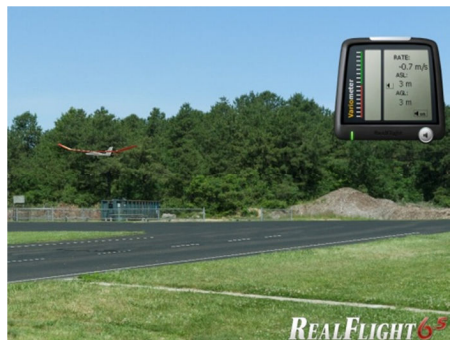 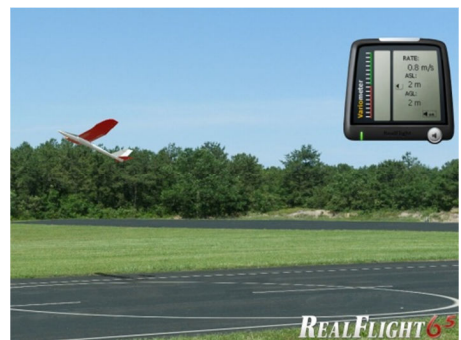

(a) Final approach    (b) Begin vertical climb

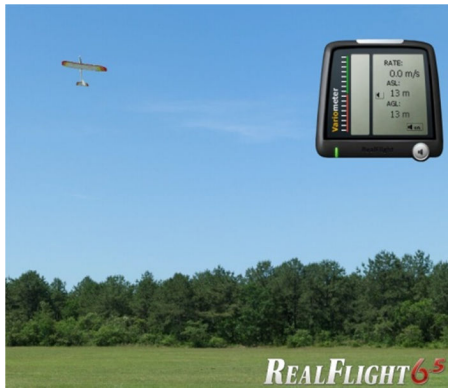 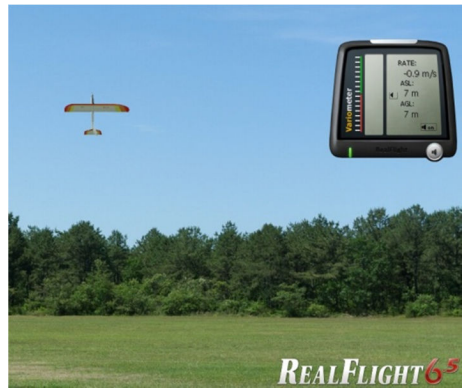

(c) Transition-state hovering

(d) Controlled descent

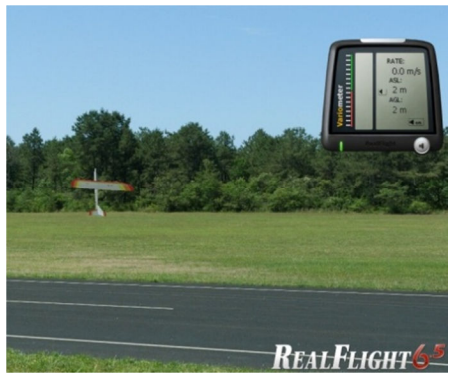 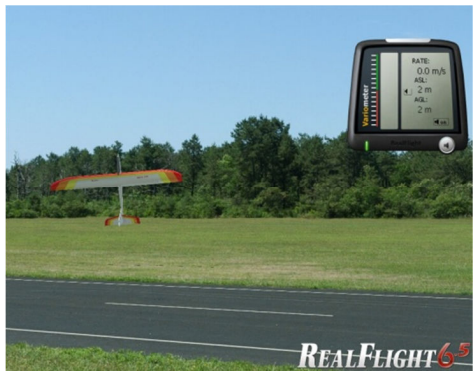

(e) Final-state hovering

(f) Precision high-alpha maneuvering onto landing station (optional)

**Fig. 4** Vertical landing sequence

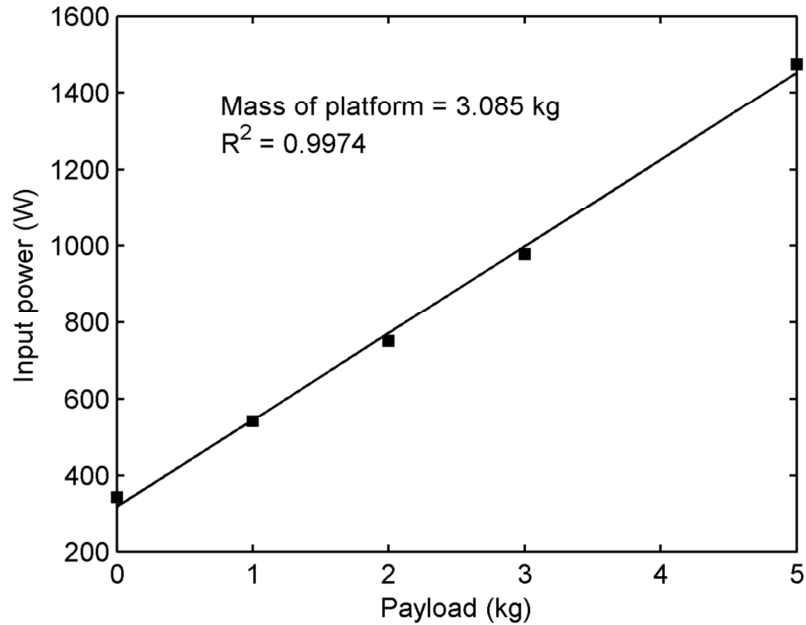

**Fig. 5** Variation of hovering power with payload

### 3.3. Gliding performance

Gliding performance of the hybrid sailplane with varying payloads was evaluated using the polar curves, as shown in Fig. 6, from which minimum sink rates and maximum glide ratios were obtained. The glide ratio refers to how far a glider can travel for a given altitude. A line is drawn from the origin tangent to the polar curve, and the speed indicated by the point of tangency is the speed to achieve maximum glide ratio [17]. The minimum sink rate is the speed at which the glider is descending at a slowest possible rate through the air [17]. Table 1 summarizes the maximum glide ratio and the minimum sink rate for different payloads, along with other crucial parameters such as the flying weights and wing loadings. It is interesting to note that despite relatively large variation in the payload, the minimum sink rate only varied between 0.6 and 0.9 ms$^{-1}$. There was little variation in the maximum glide ratio. A glide ratio of 21:1 which means that in smooth air the plane will sink 1 km for every 21 km forward. The stall speeds tabulated in

Table 1 were also derived from the polar curves (Fig. 6), and as expected, they increased with payload.

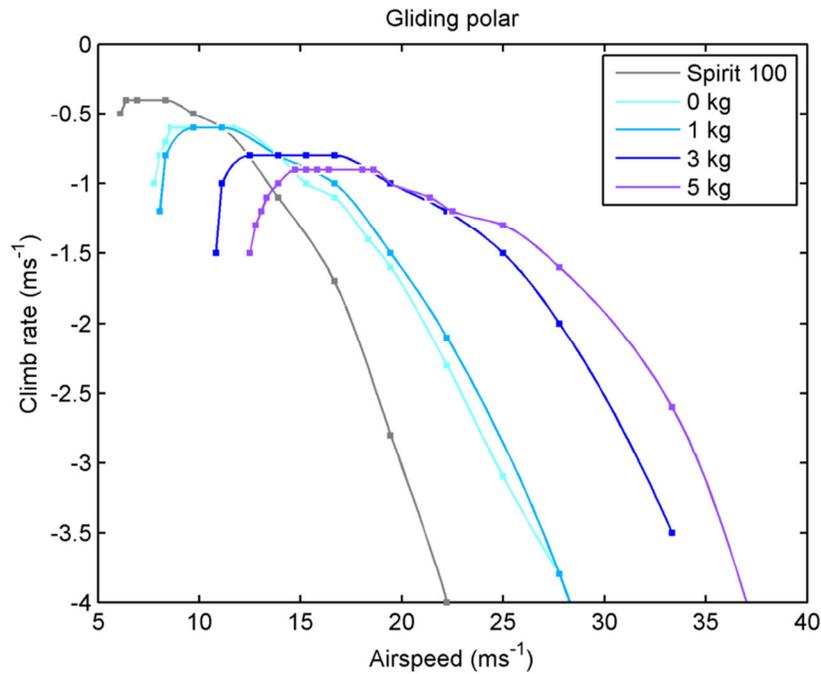

**Fig. 6** Polar curves for the original Spirit 100 and the Spirit 100-VT with different payloads.

Sink rates ($V_s$) of 0.6 to 0.9 ms$^{-1}$ found in this simulation study implied that average updraft velocity ($\bar{w}$) of 1 ms$^{-1}$ or more is sufficient to keep the platform airborne, as indicated by Eq. 1. Obviously the thermalling conditions will be more easily met with lighter payloads or a wing with a higher lift-to-drag ratio.

**Table 1** Key flight parameters for the simulated Spirit 100-VT and the effect of payload. Motorless Spirit 100 was included for comparison.

| Configuration of platform | Flying weight (kg) | Wing loading (N·m$^{-2}$) | Hovering input power (W) | Stall speed (km·h$^{-1}$) | Minimum sink rate (ms$^{-1}$) | Best glide speed (km·h$^{-1}$) | Maximum glide ratio |
|---|---|---|---|---|---|---|---|
| Spirit 100 | 1.745 | 25.27 | N/a | 22 | 0.4 | 31 | 21.1 |
| Hybrid (0 kg) | 3.085 | 44.67 | 343.68 | 28 | 0.6 | 42 | 19.7 |
| Hybrid (1 kg) | 4.085 | 59.14 | 542.82 | 29 | 0.7 | 42 | 19.0 |
| Hybrid (3 kg) | 6.085 | 88.10 | 978.8 | 39 | 0.8 | 62 | 21.1 |
| Hybrid (5 kg) | 8.085 | 117.06 | 1474.86 | 49 | 0.9 | 67 | 20.7 |

## 3.4. Analysis of gliding flight in tropical cyclone

This section evaluates the potential of Spirit 100-VT to achieve persistent flight in the eyewall of tropical cyclone by analyzing its gliding performance and the characteristics of vertical motions. From section 3.1, the mean distance between two updraft cores is 2.9 km. The maximum glide ratio of the Spirit 100-VT without payload was 19.7 (Table 1). Calculations based on its gliding performance indicated that it will lose 147.2 m in altitude while gliding to the next updraft core at its best glide speed of 11.67 ms$^{-1}$ (42 kmh$^{-1}$). From Fig. 3, the updraft cores are in close proximity to each other as the updrafts were dominant within the eyewall with the percent coverage of ~ 54% at 0.5 km altitude. Now, let's assuming there is a downdraft that exists along the gliding phase and occupying 1/4 of the path. With a top 10% average downdraft core value, $\overline{w}_{down}$ of 3.2 ms$^{-1}$ [10], this will lead to a total altitude loss of 346.1 m with remaining height of 153.9 m. Though the lowest safe altitude (LSALT) for aviation is generally 304.8 m (1000 ft), ~150 m of safety buffer should be adequate for a small UAV. In case the downdraft core value = $\overline{w}_{down\,max}$ (6 ms$^{-1}$) [10], the total altitude loss was 520.1 m which exceeded the initial flight height of 500 m. To mitigate this problem, the initial height can be increased to 600 m or powered flight can be initiated if altitude drops below 50 m. The percent area of downdraft cores in the eyewall was only 22.7% [10] and therefore the probability of encountering downdraft would not be too significant.

Fig. 7 shows the altitude loss as a function of downdraft core velocity for different payload configuration. S 100 refers to the motorless Spirit 100. As it turned out, the variant with higher wing loading suffered least altitude loss because their best glider speeds are higher while retaining glide ratio around 21:1. For that reason, it has been a common practice to equip cross-country sailplanes with water ballast to achieve higher cruising speeds [18]. These simulation results strongly suggest that persistent UAV flight in the eyewall of a tropical cyclone is feasible. Future work will extend the study to the rainbands as well as the early stages of cyclogenesis.

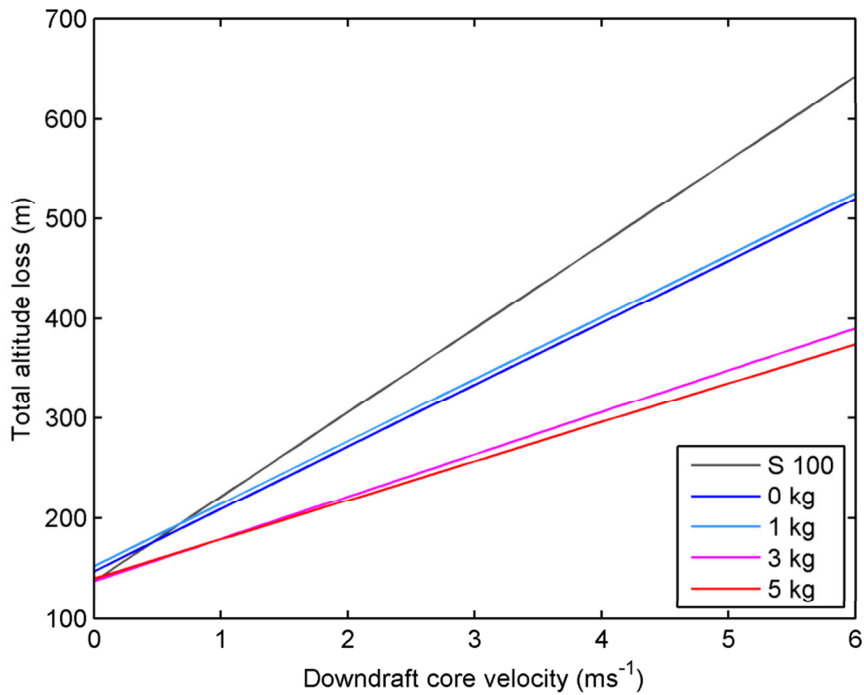

**Fig. 7** Altitude loss as a function of downdraft core velocity while gliding between two updraft cores in the eyewall region.

## Conclusions

Persistent UAV flight within the eyewall region of tropical cyclone has been proposed and simulated. Previous research and analyses of direct flight observations of mature hurricanes had found that the updraft cores covered about 55% of the eyewall regions with $\bar{w}$ of 4.5 ms$^{-1}$ at 0.5 km altitude. Given the vertical motion properties, simulation work suggested that persistent flight is indeed viable. Glide ratio of 20.7 and a wing loading of 117.06 Nm$^{-2}$ resulted in an altitude loss of under 300 m during the gliding phase across a core-to-core distance of 2.9 km. VTOL capability of the persistent UAV had also been demonstrated via simulation. Persistent flight would usher in a new era of research and observations of tropical cyclones.